\documentclass[%
 preprint,
 amsmath,amssymb,
 aps,
]{revtex4-1}

\usepackage{graphicx}% Include figure files
\usepackage{dcolumn}% Align table columns on decimal point
\usepackage{bm}% bold math
\usepackage{etoolbox}

\makeatletter
\patchcmd{\frontmatter@abstract@produce}
  {\vskip200\p@\@plus1fil
   \penalty-200\relax
   \vskip-200\p@\@plus-1fil}
  {}
  {}
  {}
\makeatother

\begin{document}

\renewcommand{\thefootnote}{\fnsymbol{footnote}}

% Author info
\title{Gel rupture in a dynamic environment}

\author{Kelsey-Ann Leslie$^a$}
 \affiliation{Department of Physics and Astronomy, Northwestern University, Evanston, IL, USA}%Lines break automatically or can be forced with \\
 
\author{Robert Doane-Solomon\footnote{these authors contributed equally to this work}}%
\affiliation{Department of Physics and Astronomy, Northwestern University, Evanston, IL, USA}
\altaffiliation{Department of Physics, University of Oxford, Oxford, UK}

\author{Srishti Arora}
 \affiliation{Department of Physics and Astronomy, Northwestern University, Evanston, IL, USA}%Lines break automatically or can be forced with \\
 
 \author{Sabrina Curley}
 \affiliation{Department of Chemical Engineering and Materials Science, Michigan State University, East Lansing, MI, USA}%Lines break automatically or can be forced with \\
 
 \author{Caroline Szczepanski}
 \affiliation{Department of Chemical Engineering and Materials Science, Michigan State University, East Lansing, MI, USA}%Lines break automatically or can be forced with \\

\author{Michelle M. Driscoll}
 \affiliation{Department of Physics and Astronomy, Northwestern University, Evanston, IL, USA}%Lines break automatically or can be forced with \\

%\date{}% It is always \today, today,

\begin{abstract}
Hydrogels have had a profound impact in the fields of tissue engineering, drug delivery, and materials science as a whole. Due to the network architecture of these materials, imbibement with water often results in uniform swelling and isotropic expansion which scales with the degree of cross-linking. However, the development of internal stresses during swelling can have dramatic consequences, leading to surface instabilities as well as rupture or bursting events. To better understand hydrogel behavior, macroscopic mechanical characterization techniques (e.g.\ tensile testing, rheometry) are often used, however most commonly these techniques are employed on samples that are in two distinct states: (1) unswollen and without any solvent, or (2) in an equilibrium swelling state where the maximum amount of water has been imbibed. Rarely is the dynamic process of swelling studied, especially in samples where rupture or failure events are observed. To address this gap, here we focus on rupture events in poly(ethylene glycol)-based networks that occur in response to swelling with water. Rupture events were visualized using high-speed imaging, and the influence of swelling on material properties was characterized using dynamic mechanical analysis. We find that rupture events follow a three-stage process that includes a waiting period, a slow fracture period, and a final stage in which a rapid increase in the velocity of crack propagation is observed. We describe this fracture behavior based on changes in material properties that occur during swelling, and highlight how this rupture behavior can be controlled by straight-forward modifications to the hydrogel network structure.
\end{abstract}

\maketitle

\section{Introduction}

Polymer materials formed with a significant degree of cross-linking, thus creating a network structure, have specific and unique advantages when compared to linear analogues that have no cross-linking.~\cite{Creton2017}
Most significantly, network architectures are associated with additional rigidity and stability of three-dimensional shape and structure. These changes correspond to an increase in mechanical integrity, typically measured by the elastic modulus, as well as an enhanced resistance to thermal degradation and dissolution in solvents. Over the past 20 years, hydrophilic polymer networks that swell but do not dissolve when immersed in water, e.g.\ hydrogels, have been exploited heavily as platforms for biomaterials and synthetic mimics of tissues, most commonly with poly(ethylene glycol) (PEG) derivatives employed as the polymer network building block.~\cite{ZHU20104639, Okay2009} Hydrogels have proven valuable for these applications, as they can imbibe a significant amount of water, thus generating a soft material that is mostly comprised of water yet has a defined structure dictated by the polymer network.  While the research literature is rich with examples of biomaterials developed from PEG,~\cite{Nemir2010,ZHU20104639,Shoichet2010,caliari2016,galarraga2019} there are still challenges associated with this platform: there remains a need to better understand the physical behavior and transient dynamics associated with swelling and how it contributes to observed phenomena such as surface instabilities and rupturing.~\cite{Tanaka1979,deSilva2016,ILSENG2019,Wang2019,Curatolo2017,Takahashi2016,liu2015advances} To address these limitations, here we employ high-speed imaging of rupture events during hydrogel swelling to determine what factors lead to catastrophic failure.
    
A major drawback of hydrogel platforms is their associated brittle behavior and overall lack of toughness,~\cite{Imran2010,Guo2020} which limits their use in emerging applications. These drawbacks impact hydrogels formed from free radical polymerization of monomer solutions, including rapid, on-demand photopolymerizations of acrylate- or methacrylate- PEG derivatives. Hydrogels formed from these precursors typically have heterogeneity in their network architectures~\cite{KANNURPATTI1998} and display failure responses that are highly subjective to the type of applied load or stress.~\cite{Chester2012,CHENG2019} A variety of responses including fast fracture, delayed fracture, fatigue fracture, and catastrophic fracture~\cite{TANG201724} have all been documented, yet little work has been done to identify a fundamental relationship that governs these different mechanisms.  Despite this gap, the ease of applying free-radical polymerizations and more specifically the advantages associated with photo-initiated free-radical polymerizations (e.g. spatio-temporal control, ambient temperature processing~\cite{Chatani2014}) makes it a heavily-relied upon approach that is employed across numerous application fields. Therefore, this approach is necessary for future innovations and developments such as complex and patterned hydrogels,~\cite{Cativa2019,Nemir2010,Bryant2006} but the mechanics of these materials must be understood.

One specific shortcoming when assessing failure events or instabilities of gels is that they are typically studied in one of two distinct and separate states: (1) unswollen, i.e.\ in a virgin state without any solvent; or (2) at an equilibrium swelling state. In the unswollen regime, the network is in a collapsed state and there are no polymer/solvent interactions.  In contrast, equilibrium swelling assumes the material is in thermodynamic equilibrium with a solvent that has an affinity for the polymer network; it is commonly assumed that the swollen network is in a stress-free state.~\cite{BOSNJAK2020,Curatolo2017} The ability of a network to expand isotropically and be impregnated with an attractive solvent is correlated to the relative degree of cross-linking, and thus more densely cross-linked networks swell to a lesser extent than loosely cross-linked analogues. Characterization of polymer gels in just these two distinct and separate states (swollen and unswollen) can be sufficient for certain application purposes, as well as to understand the physics of these two endmember regimes. Prior works have used these distinct states to determine how boundary conditions during swelling~\cite{Chester2012} or solvent cycling (i.e.\ repeated swelling and drying)~\cite{BOSNJAK2020} impact mechanical behavior, revealing that in an unswollen state, standard viscoelastic behavior associated with polymer networks is evident, whereas at equilibrium swelling the gel behaves as a non-linear elastic material.~\cite{BOSNJAK2020} Unfortunately, while these studies provide critical insight into the behavior of gel materials, they do not capture the mechanics in dynamic or evolving environments, particularly during the swelling process, nor do they account for any failure events or instabilities encountered as a direct consequence of swelling.

The mechanical behavior of swelling has been studied from a theoretical perspective, initially through the development of Flory-Rehner model,~\cite{Flory1982} which considered the free energies associated both with polymer stretching as well as the mixing of polymer and solvent (as previously derived in the Flory-Huggins solution theory~\cite{Flory1941,Huggins1941}). Since this foundational work, others including Tanaka and Fillmore have combined theoretical considerations with experimental analyses to understand the swelling of gel materials.~\cite{Tanaka1979} In Tanaka and Fillmore's analysis, it is assumed that positive osmotic pressure forces a gel to expand when transferred to a compatible fluid, and that the speed of swelling is proportional the the diffusion coefficient (\textit{D}) of the polymer network. Furthermore, while approaching the equilibrium swelling state the internal stresses of the polymer network approach zero. This was shown to be a reasonable assumption for polyacrylamide gels in water having spherical shapes.~\cite{Tanaka1979} However, since this analysis more recent studies have revealed surface instabilities~\cite{Tanaka1987} and self-rupturing~\cite{deSilva2016} behavior in response to swelling, and thus the appropriateness of these theoretical regimes to account for geometrical constraints, mechanical loads, and non-idealities or heterogeneity associated with polymer networks used in practice has come into question.~\cite{OKUMURA201661} Phenomena such as \emph{in situ} surface structuring and self-rupturing are interesting and relevant for interfacial design and drug delivery applications. Unfortunately, given the prior analyses of failure modes and mechanisms only in static states of swollen or unswollen, limited knowledge exists regarding the failure events of polymer gels in response to internal stresses developing in real-time. Furthermore, while growing interest has led to theoretical analyses that take into account the swelling process,~\cite{Tanaka1979,COHEN2019666,Chester2012} an experimentally-driven approach to characterizing and understanding swelling-induced instabilities and failures will better inform these models and provide \textit{ab initio} predictions of failure events in gels.

In this work, we demonstrate the necessity of probing gel properties in dynamic environments, e.g.\ during the swelling process, and not just in the swollen and unswollen states.  We study in particular how fluid imbibement leads to material failure, and how this rupture is controlled both by dynamically changing material properties, as well as the imbibement-induced stress that builds up due to osmotic pressure gradients and inhomogeneous material properties.  Our measurements  demonstrate that both material properties and stress increases due to fluid flow control the crack velocity in these gels.  This work highlights the rich spectrum of behavior that can occur during the swelling process, and we expect this to motivate further investigations of this under-explored regime.

\section{Materials and Methods}
\subsection{Materials}
All experiments were conducted using photopolymerized polymer networks.  Networks were formed from poly(ethylene glycol) (PEG) monomers, specifically: poly(ethylene glycol) methyl ether acrylate (M\textsubscript{n} $\approx$ 480, PEGMEA) and poly(ethylene glycol) diacrylate (M\textsubscript{n} $\approx$ 700, PEGDA).  A photoinitiator was included in all monomer formulations at 0.5 wt\% (relative to the total mass of monomers) to enable on-demand photopolymerization. In all cases, the photoinitiator employed was 2,2-dimethoxyphenylacetophenone (DMPA). All monomers and the photoinitiator were purchased from Sigma-Aldrich and used as received and without further purification. For all swelling experiments Milli-Q water was employed.

\subsection{Methods}
\subsubsection*{Preparation of Monomer Formulations}
The ratio of PEGDA:PEGMEA was varied to systematically modify the  mechanical properties of the PEG-networks, similar to approaches described in prior works.~\cite{Okay2009, Hoshino2018} The mol percent of PEGDA was varied over a range of 1 - 40\%; a formulation labeled 20\% PEGDA indicates that of the monomers included in the formulation, 20 mol percent is the cross-linker, PEGDA, and the remaining 80 mol percent is PEGMEA. To prepare monomer formulations, a given  ratio of PEGDA:PEGMEA was determined and appropriate masses of each monomer were placed into a clean glass vial. The appropriate mass of photoinitiator (DMPA) corresponding to a 0.5 \% (w/w) loading level, was introduced and magnetic stirring and/or vortexing was applied until a homogeneous, clear solution was obtained upon visual inspection (typically after 5-10 minutes of mixing). If not employed immediately, monomer formulations were placed in a refrigerator and covered in aluminum foil to prevent premature UV exposure during storage. 

\subsubsection*{Photopolymerization of Unswollen Specimens}
 For samples that were monitored in real-time for fracture and rupture events, liquid PEGDA:PEGMEA monomer formulations were deposited into the cavity of 
 a silicone square mold (internal dimensions: 3 cm $\times$ 3 cm $\times$  0.3 cm) 
and photopolymerized using a UV LED (ThorLabs, Solis-365A, I\textsubscript{o} = 50 mW/cm\textsuperscript{2}) for approximately 10 seconds. To polymerize specimens intended for mechanical characterization via dynamic mechanical analysis (DMA), a slightly modified procedure was employed with I\textsubscript{o} = 250 mW/cm\textsuperscript{2} (as measured with an EIT radiometer) and samples polymerized for 420 s. Furthermore, samples were cured using glass slides as spacers to create bar-shaped specimens appropriate for DMA characterization and with approximate dimensions of 25 mm $\times$ 10 mm $\times$ 2 mm (l $\times$ w $\times$ h).
 
\subsubsection*{Transient Swelling Characterization}
To characterize the transient swelling behavior of networks with varying PEGDA:PEGMEA fractions, the sample mass was first measured in the dry, unswollen state. Immediately after recording this mass, samples were immersed in 500 mL of Milli-Q water (at ambient temperature, T = 27$^{\circ}$C). At pre-defined measurement times (typically every 30s), specimens were removed from the water and the elapsed swelling time was paused. Specimens were gently dried with a Kimwipe to ensure that no excess water was included in the measurements, and then weighed using the same balance as initial mass measurements. After weighing, specimens were then placed in a fresh batch of Milli-Q water and the elapsed swelling time was re-initiated. The swelling ratio was calculated at each timepoint as the ratio of swollen mass at time \textit{t} to initial, dry mass.\cite{Okay2009} 

\subsubsection*{Fracture Observation}
To ensure no interaction with the container boundary, specimens were placed in a large glass beaker of Milli-Q water (the beaker diameter was several times larger than the size of the sample).  Fracture events were imaged from above using a high speed camera (Phantom VEO 640L); all movies were filmed at 3000 frames per second. The gel samples were backlight using a broadband LED light (Phlox Corp, White LED Backlight 100x100) to enable visualization of fracture events. A static image of a linear scale (ruler) was taken for each movie, and then was used to set the image scale during analysis. 

\subsubsection*{Calculation of Crack Propagation Velocity}
All image analysis was performed using ImageJ. Crack velocity was measured from the first moment a visual difference in a crack was observed between two frames of the high-speed videos collected; the distance the crack propagated was measured for each collected video frame. Therefore, crack velocity could be measured as \textit{d/t}, where \textit{d} represents the distance traveled by the crack in one frame and \textit{t} is the elapsed time associated with each frame. 

\subsubsection*{Bulk Network Mechanical Characterization}
Thermomechanical properties, mainly the storage modulus ($E'$) and loss modulus ($E''$), were measured using a dynamic mechanical analyzer (DMA, TA 850, TA Instruments). Specimens were analyzed using a temperature sweep from ambient conditions, 24$^{\circ}$C, to 100$^{\circ}$C at heating rate of $3^{\circ}$C/min, with 0.01\% strain and a frequency of 1 Hz. $E'$ data was analyzed with SciDavis. Thermomechanical data was collected for both unswollen and swollen samples. For unswollen samples, analysis proceeded directly as described above. For swollen specimens, samples were massed prior to soaking in distilled water for a period of 30, 60, 120, 240 or 300 s. After the soaking period, samples were massed again (as done with the swelling characterization) to determine the amount of water uptake. Samples were then immediately placed in the DMA for thermomechanical characterization. After the DMA run, samples were then massed a final time to determine the impact of the temperature sweep on degree of swelling, as discussed in the Results and Discussion.

\section{Results/Discussion}

\subsection{Mechanical properties}

\begin{figure*}[h]
\centering
  \includegraphics[width = 0.95\columnwidth]{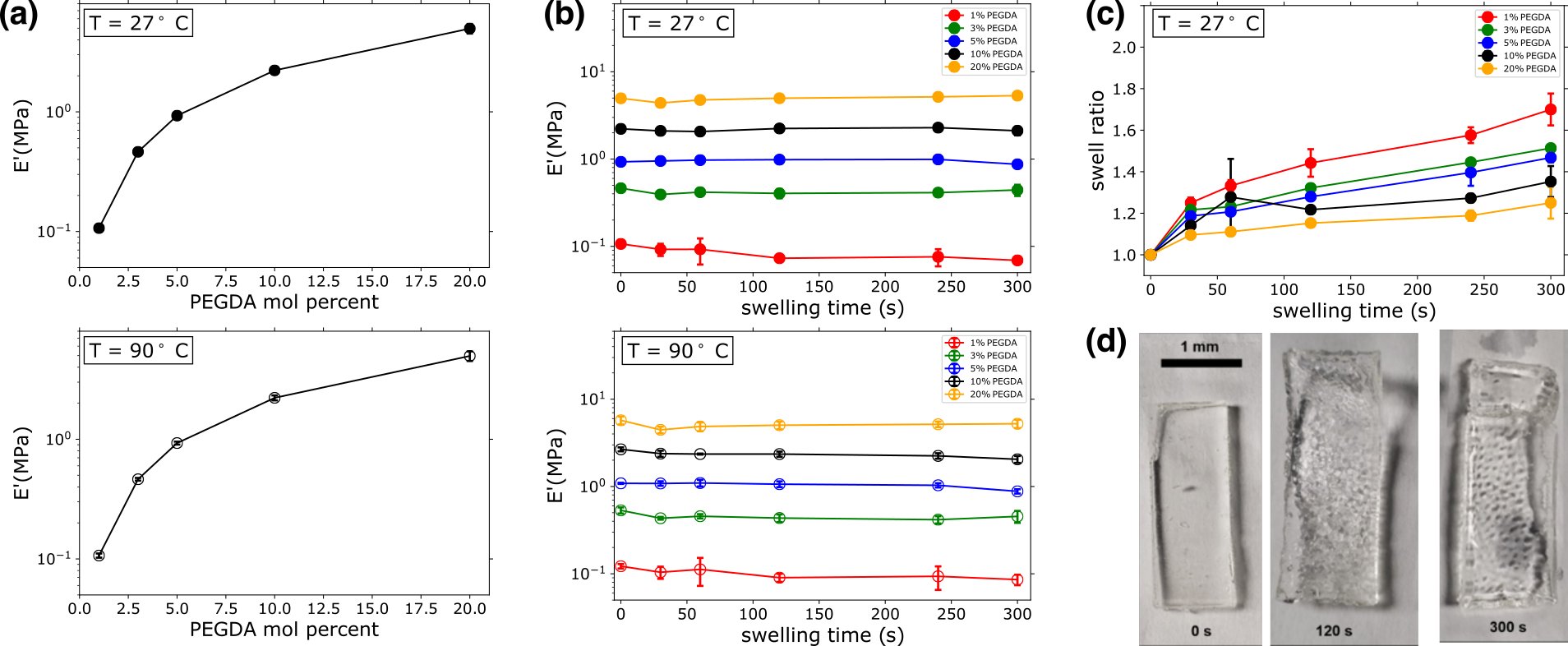}
\caption{(a) Gel storage modulus vs.\ PEGDA mol percent at T = 27$^\circ$ C (upper panel) and T = 90$^\circ$ C (lower panel). (b) Gel storage modulus as a function of swelling time at T = 27$^\circ$ C (upper panel) and T = 90$^\circ$ C (lower panel).  (c) Swelling ratio at T = 27$^\circ$ C. (d) Images of a PEG-hydrogel (3:97 PEGDA:PEGMEA) during the swelling process; the last panel shows the ruptured gel.}
\label{mech_properties}
\end{figure*}

PEGDA:PEGMEA networks were formed via bulk photopolymerization (i.e.\ with no solvent present), and afterwards probed for mechanical properties such as tensile storage modulus ($E'$) via dynamic mechanical analysis (DMA). As expected, $E'$ was directly set by the ratio of PEGDA:PEGMEA. As shown in Figure \ref{mech_properties}a, systematic increases in PEGDA fraction result in an increased value of $E'$, with an order of magnitude increase observed as the amount of PEGDA was varied from 3-20 mol percent. The measurement of $E'$ provides an estimate of stored energy within a viscoelastic material, and confirms the expected increases in rigidity and mechanical integrity with increasing PEGDA mol percent in the unswollen state. This increase in $E'$ with PEGDA mol percentage is consistent at both ambient (27 $^{\circ}$C) and elevated temperatures (90 $^{\circ}$C), as shown in Figure \ref{mech_properties}a. The gel mechanical properties we measure are additionally confirmed by our observations: all samples had a gel-like consistency when photopolymerized in bulk, indicating rubbery behavior even at ambient temperatures. Other thermomechanical data (not shown) was collected during these experiments, such as the loss modulus, $E''$, and tan $\delta$. Over the temperature range explored, $E''$ was stable and consistently three orders of magnitude lower than $E'$ for all samples. Viscoelastic materials are expected to have significant variations over orders of magnitude in both storage ($E'$) and loss ($E''$) modulii when going through the glass transition, followed by a rubbery plateau region characterized by relatively consistent moduli values above the glass transition temperature ($T_g$). The consistent modulii values at both 27 $^{\circ}$C and 90 $^{\circ}$C observed here, as well as the qualitative observations noted above, indicate that all samples are indeed in the rubbery regime of the viscoelastic profile at ambient temperatures or higher, meaning the $T_g$ of all samples is well below ambient conditions.  This is consistent with reported literature values for PEGDA-based networks of $T_g \sim -45^{\circ}$C.\cite{Kalika2006,Keim2010}

DMA was also employed to determine the impact of transient swelling and how $E'$ varied as samples were swollen prior to any macroscopically observed fracture or failure events, as illustrated in Figure \ref{mech_properties}b. Samples were loaded immediately after the swelling period, and a temperature sweep was initiated from 24$^{\circ}$C. Therefore, the $E'$ data in the upper panel of \ref{mech_properties}b is an estimate of the modulus in the swollen state. No significant variations in $E'$ were observed across swelling times up to 300 s. With the least cross-linked materials (1 mol percent PEGDA), a slight variation in $E'$ is observed between 0 s swelling (i.e.\ unswollen) and 30s. While the observed variation is quite small, it is attributed to the increased swellability of the 1 mol percent PEGDA sample due to its low cross-link density. We massed all samples after thermomechanical characterization (i.e.\ after having completed the temperature sweep from 24$^{\circ}$C to 100$^{\circ}$C), and it was determined that water loss resulted from the temperature variation and environment inside the DMA chamber, bringing the samples back to their original mass prior to swelling. Therefore, the $E'$ data at 90$^{\circ}$C (Figure \ref{mech_properties}b, lower panel) and the fact that is nearly identical to the measurements in the swollen state at 27$^{\circ}$C  highlight two significant points. First, no significant breakdown of the network structure occurs in the early stages of swelling, as the network maintains similar $E'$ values. If damage to the network architecture was occurring in the early stages of swelling one would expect a reduction in $E'$. Secondly, the fact that similar moduli values were recorded for swollen specimens (i.e.\ the $E'$ measurements at 27$^{\circ}$C) and specimens that were  swollen and dried (i.e.\ the $E'$ measurements at 90$^{\circ}$C) highlights that swelling does not have a major impact on $E'$. The water uptake was measured for all DMA specimens, and it scaled with the PEGDA mol percent and increased with swelling time, as expected (see Figure \ref{mech_properties}c); Figure \ref{mech_properties}d shows images of the gel during the swelling process. A maximum swelling ratio of $\sim1.7$  was measured after 300 s swelling for 1 mol percent PEGDA samples, whereas the most cross-linked sample (20 mol percent PEGDA) had a swelling ratio of $\sim1.2$ after 300 s. Significant decreases in $E'$ with swelling have been reported in other studies of hydrogels, however this is often observed in samples swelling on the scale of hours or days and with swelling ratios significantly larger than those measured here (on the order of 10-100).~\cite{Subramani2020}  Due to these relatively small swelling ratios, we can conclude that the fluctuations in $E'$ are minimal and only visually observed in the ``softest'' or least cross-linked sample.

\subsection{Dynamic fracture}

\begin{figure*}[h]
\centering
  \includegraphics[width=0.95\columnwidth]{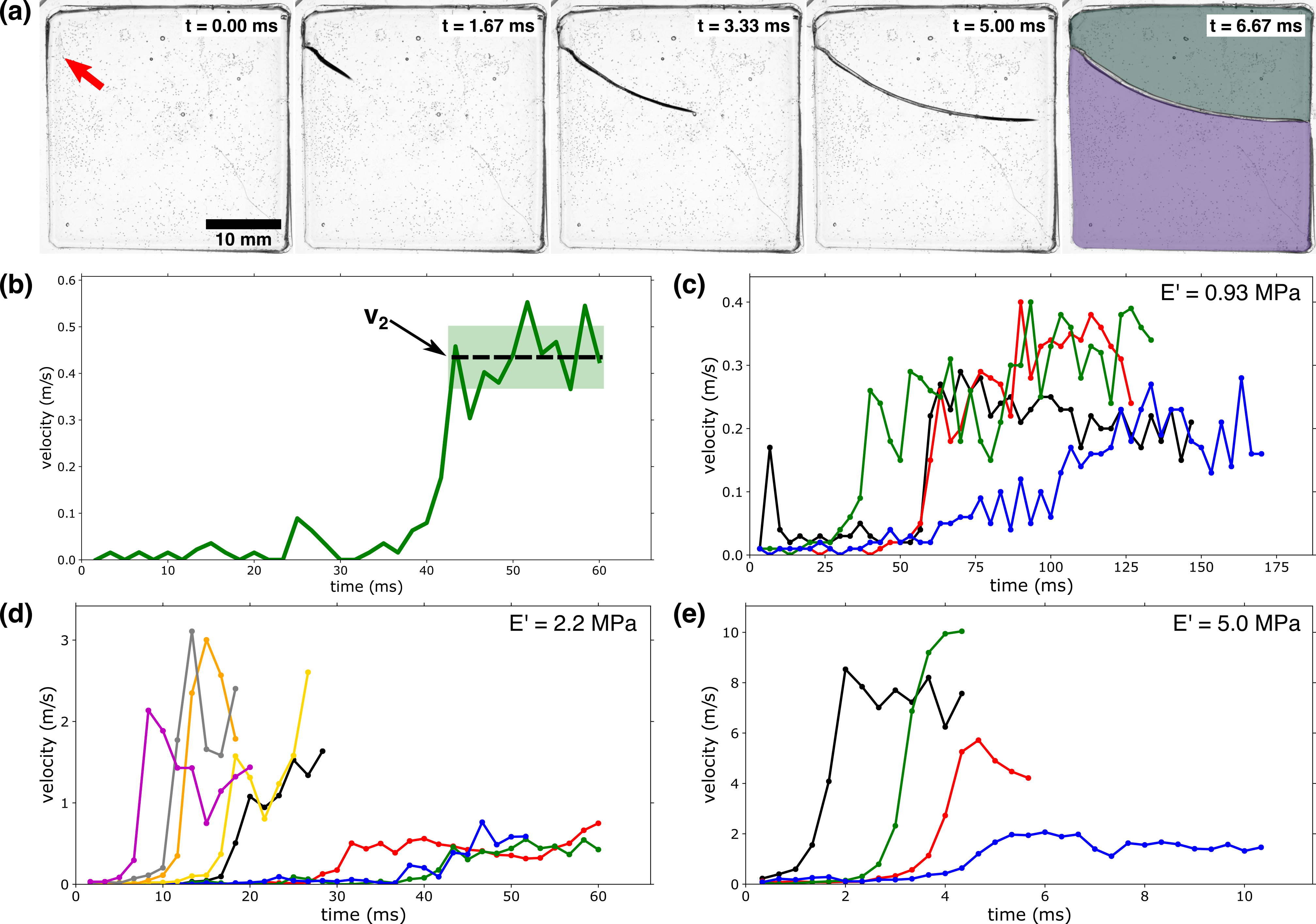}
\caption{\textbf{Crack velocity profiles:} %(a) velocity_figure3: images are from video 20A. 
(a) A high-speed camera was used to capture a PEG-hydrogel (E' = 5.0 MPa) rupturing violently due to fluid-induced compressive stress. The red arrow in the first panel indicates the initiation point of the running crack. In the final stage, the gel has broken into two separate pieces as indicated by the colored overlays. (b) Identification of $v_2$: as indicated by the dashed line, $v_2$ is demarcated by a large velocity jump followed by a near-constant velocity region. The shaded area indicates the fluctuations in $v_2$; these fluctuations are used to calculate the error bars in Figure \ref{v2}a. (c-e) Crack velocity profiles for (c) E' = 0.93 MPa , (d) E' = 2.2 MPa, and (e) E' = 5.0 MPa. Only the first 80\% of growth is shown to avoid edge interactions.~\cite{Goldman2010}}
\label{velocity_profiles}
\end{figure*}

When immersed in water, PEGDA:PEGMEA gels begin to absorb fluid.  This swelling occurs at a slow rate, resulting in a buildup of compressive stresses at the gel boundary.~\cite{Curatolo2017} We observed that this swelling occurred with no damage to the gel samples for a time period $\tau$; after this time, the samples fractured; one such event is illustrated in Figure \ref{velocity_profiles}a. In the softest gels that fractured, crack branching was observed. 
This branching became more and more suppressed as the gel modulus was increased (increasing PEGDA mol percent); for the stiffest gels crack branching was quite rare.

Fracture behavior was only observed for a limited range of gel moduli, 0.5-5 MPa (3-20 mol percent PEGDA).  Softer samples, with moduli below 0.5 MPa (3 mol percent PEGDA)  were observed to uniformly swell without fracturing. Samples with moduli above 5 MPa (20 mol percent PEGDA) exhibited little swelling and remained intact, though rare fracture events were observed. These two extremes can be explained by considering the viscoelasticity associated with PEG-networks. Below 3 mol percent PEGDA, the low degree of cross-linking increases the contribution of the viscous portion of the complex mechanical behavior of the network and $E'$ is reduced (see Figure \ref{mech_properties}a). Therefore, dissipation of applied stresses and internal relaxations can occur without damage to the network, meaning rupture events or instabilities are not observed. At the other extreme, above 20 mol percent PEGDA, the cross-link density and associated $E'$ are so high  that swelling is severely limited and thus no significant stresses build up internally; this eliminates the instabilities that result from swelling.

The rupture of our hydrogel samples was observed to occur as a  three-stage process: first the gel imbibes fluid for a `waiting time', $\tau$, with no visible rupture; second, fracture  begins as a slow, creep-like process; finally, the rupture velocity rapidly increases to a new higher velocity, $v_2$.  The waiting time, $\tau$, occurs on a timescale of hundreds of seconds, while the two-stage rupturing occurs on a time scale of ms; we note that due to the large difference in these timescales, the velocity vs.\ time plots in Figure \ref{velocity_profiles}c-e do not indicate the waiting time.  First, we will consider the final two stages of the rupture process where the crack is actively expanding; Figure \ref{velocity_profiles}b illustrates how we define the final crack velocity $v_2$.  As shown in Figure \ref{velocity_profiles}c-e, we observed large fluctuations from sample to sample in both the value of $v_2$ as well as the initiation time of the third stage of the rupture process.

To understand the complex rupture process that occurs in our hydrogels, we must first consider how rupture proceeds in a more standard elastic material.  All materials are embedded with microscopic cracks, voids, or other flaws; it is the growth of these `micro-cracks' that causes brittle materials to crack into pieces under an applied load.  As first derived by Griffith in 1921,~\cite{griffith1921phenomena} and later refined by many others,~\cite{fineberg1999instability, Freund1998} an energy balance argument can be used to determine $l_c$, the maximum length of a stable microcrack as a function of the boundary stress, $\sigma$, elastic modulus, $Y$, and the energetic cost of creating new surfaces, $\gamma$:

\begin{equation}
\label{griffiths}
    l_c \sim \gamma Y / \sigma^2.
\end{equation}

Cracks that are smaller than $l_c$ are stable, while those longer than this are not; it is energetically favorable for cracks larger than $l_c$ to spontaneously elongate (We note that equation \ref{griffiths} may contain a pre-factor that is determined by the precise loading and fracture geometry).  The first stage of the gel rupture process is exactly a reflection of this mechanism: the hydrogel samples experience compressive stress when they are immersed in water due to the osmotic pressure gradient; this stress increases as the gel imbibes more fluid and its elastic properties become highly non uniform.  Eventually, these stresses become large enough to initiate rupture via expansion of a crack.

Once a crack begins to run in a brittle material, fracture mechanics predicts that it will accelerate until it is growing at the Rayleigh velocity, $c_R$~\cite{Freund1998}:

\begin{equation}
\label{rayleigh}
    c_R = f(\nu)\sqrt{G/\rho},
\end{equation}

where $G$ is the material's shear modulus, $\rho$ is the material's density, and $f(\nu)$ is a function of the material's Poisson ratio, and varies between 0.87-0.95. It should be noted that instabilities such as oscillations and microbranching often occur, which limit the maximum crack speed to a fraction of $c_R$ (typically 0.3-0.7)~\cite{fineberg1999instability,Freund1998}.  As discussed previously, we observe a two-stage expansion process.  We are uncertain of the origin of the initially slower growth, though the pattern of slow acceleration to limiting crack velocity has been seen in other experiments on soft, brittle materials.~\cite{livne2005universality, deegan2001oscillating}

\begin{figure}[h]
\centering
  \includegraphics[width = 0.42\columnwidth]{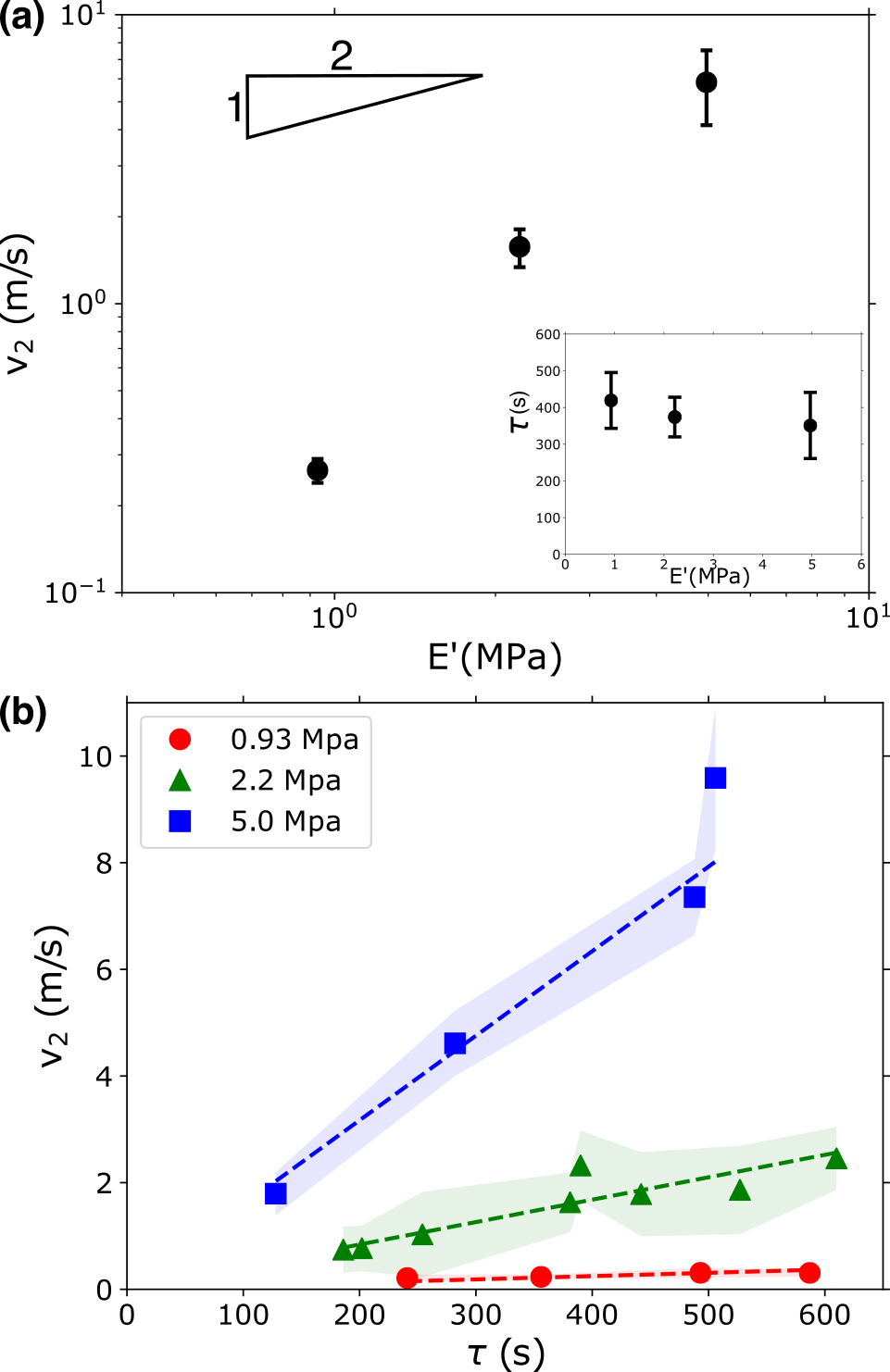}
\caption{(a) The mean crack velocity, $v_2$, increases with increasing gel modulus. However, $v_2$ does not follow the square root scaling expected for a brittle material (square root slope is indicated by the inset triangle). The lower right inset shows that the waiting time, $\tau$, is uncorrelated with gel modulus.  (b)  $v_2$ increases linearly with increasing fracture start time, $\tau$; the fluctuations in $v_2$ are indicated by the shaded region. }
\label{v2}
%\vspace{-7mm}
\end{figure}

What sets the value of $v_2$?  Equation \ref{rayleigh} suggests that the physical parameters that set the value of $v_2$ are the gel modulus and density.  Fig \ref{v2}a shows that $v_2$ indeed does increase as the gel becomes stiffer, but this increase does not follow the square root dependence predicted by Equation \ref{rayleigh}.    This scaling can be understood by considering the uptake of fluid into the gel that occurs during  the waiting time $\tau$. The gel swells and imbibes fluid; this does not have an appreciable effect on the gel's modulus (see Figure \ref{mech_properties}b), but the gel does decrease in density as it absorbs water.  This occurs because the polymer network expands as it imbibes fluid, increasing the gel's volume; additionally, the bulk PEGDA:PEGMEA networks in the unswollen state and immediately after photopolymerization are all more dense than water, so the water absorbed by the network reduces the overall density as the material transitions from an unswollen gel to a hydrogel. These combined effects cause a reduction in density of 10-20\% on the timescales associated with the waiting time before fracture occurs, $\tau$.
We note that the waiting time fluctuated from sample to sample, and is uncorrelated with gel modulus, see inset in Figure \ref{v2}a.  However, we found that $\tau$ was strongly correlated with $v_2$.   As shown in Figure \ref{v2}b, $v_2$ increases linearly with $\tau$.  Though the functional form is incorrect, this behavior is consistent with Equation \ref{rayleigh}: a larger $\tau$ implies more imbibement, and thus a lower density, and therefore a larger crack velocity.

The dynamic imbibement process studied here is highly complex, resulting in both continuously changing as well as non-uniform material properties, which are potentially competing effects.  This complexity is reflected in the atypical fracture behavior we observe, even though (static) unswollen hydrogels are often considered to behave as traditional brittle materials.

\section{Conclusions}
Previous studies on PEG-hydrogels have largely focused on characterizing these materials in one of their two equilibrium states: unswollen or swollen.  Here, we have examined these materials \emph{during} the swelling process, with a particular focus on characterizing the dynamics of swelling-induced rupture.

As these materials begin to imbibe water, they swell.  In highly cross-linked gels (above 20 mol percent PEGDA), this swelling is limited, and little excess stress is built up.  In loosely cross-linked gels (below 3 mol percent PEGDA), a large amount of swelling occurs, but this does not lead to a build-up of excess stress, as it is mitigated via viscous dissipation.  In between these two extremes however, imbibement leads to accumulation of enough excess stress to cause material failure.  This failure is markedly different that what is observed in standard brittle materials, reflecting the complex dynamics that occurs during the swelling process.  We observe that fracture occurs as a three-stage process, that includes a waiting period, a slow fracture period, and a final stage in which a rapid increase in the velocity of crack propagation is observed.  We find that $v_2$, the crack velocity during the third stage, is controlled by both the material's elastic modulus, and additionally by the duration of the waiting period, which reflects the large changes in material properties that occur during the imbibement and swelling processes.

This works highlights the diversity of behaviors that occur during the non-equilibrium swelling process.  While we have focused on rupture here, other works have demonstrated that a variety of elastic instabilities occur during the dynamical swelling process.~\cite{Tanaka1987}  Using these results as a starting point, further studies can illuminate how to control or prevent these instabilities and rupture processes, perhaps via material patterning or by altering the nature of the material cross-links.

\begin{acknowledgments}
During this study, SJC was supported by the College of Engineering Distinguished Scholar Fellowship from Michigan State University. 
\end{acknowledgments}

%\nocite{*}

\bibliography{gel_bib}% Produces the bibliography via BibTeX.

\end{document}